\documentclass{eptcs}
\providecommand{\event}{SOS 2007} 
\usepackage{breakurl}             
\usepackage{amssymb}
\usepackage{lscape}
\usepackage{graphicx} 
\usepackage{hyperref}
\usepackage{url}
\usepackage{multirow}
\usepackage{listings}
\usepackage{AMMALanguages}

\title{Analyzing Flowgraphs with ATL}
\author{Valerio Cosentino \and Massimo Tisi \and Fabian B\"uttner
\institute{AtlanMod, INRIA \& \'{E}cole des Mines de Nantes, France}
\email{\{valerio.cosentino, massimo.tisi, fabian.b\"uttner\}@inria.fr}
}
\def\titlerunning{Analyzing Flowgraphs with ATL}
\def\authorrunning{V. Cosentino, M. Tisi \& F. B\"uttner}
\begin{document}
\maketitle

%
%
%

\begin{abstract}
This paper presents a solution to the Flowgraphs case study for the Transformation Tool Contest 2013 (TTC 2013). Starting from Java source code, we execute a chain of model transformations to derive a simplified model of the program, its control flow graph and its data flow graph. Finally we develop a model transformation that validates the program flow by comparing it with a set of flow specifications written in a domain specific language. The proposed solution has been implemented using ATL.
\end{abstract}

\section{Introduction}

This paper presents an ATL-based solution to the Flowgraph Case Study\cite{Horn13} for the Transformation Tool Contest 2013 (TTC 2013)\footnote{\url{http://tinyurl.com/TTC2013-HomePage}}. The main task of the case study is deriving the program dependence graph (PDG) of the given source code. This graph contains both control and data flow information and is obtained through a sequence of steps: 1) creation of a simplified model for the Java program; 2) generation of the program control flow graph; 3) addition of data flow dependencies to create the PDG. A final additional task is 4) validation of the resulting PDG against a set of specifications written in the provided DSL.

The solution\footnote{The full solution is available on the SHARE server of the contest \url{http://tinyurl.com/ATL-solution}} is implemented using an ATL transformation chain. We address all the tasks of the case study by relying exclusively on the ATL declarative language, with the exception of text-to-model injectors and global orchestration. The case study shows the flexibility of ATL in handling a wide range of tasks: classical model-to-model transformation and model-to-text transformation in task 1, in-place refinement in task 2, a complex algorithm in task 3, model validation in task 4. It is also intended as a full-range example for new ATL developers.

The ATL Transformation Language (ATL) \cite{ATL} is a model transformation
language and tool available from the Eclipse modeling project \footnote{\url{http://www.eclipse.org/m2m/atl/}}. ATL is a declarative language
allowing the specification of transformation rules, that are matched over the source model
to create elements in the target model. Expressions are written using the Object Constraint Language (OCL \footnote{\url{http://www.omg.org/spec/OCL/2.3.1/}}). ATL contains also an imperative part allowing
to handle cases whose declarative expressions would be too complex. The solution we propose in this paper makes use only of the declarative part of the language.

ATL allows the developer to decorate the input metamodel with derived attributes and operations on model elements, named \emph{Helpers} and grouped into reusable \emph{Libraries}. Finally the developed transformation can be applied in \emph{normal mode}, where target models are built from scratch, or in \emph{refining mode}, where the input model is modified in-place.

This paper is structured as follows: Section \ref{sec:solution} illustrates our solution and finally Section \ref{sec:conclusions} concludes the paper.

\section{An ATL solution} \label{sec:solution}

\begin{figure}[t]
	\center
	\includegraphics[width=\textwidth]{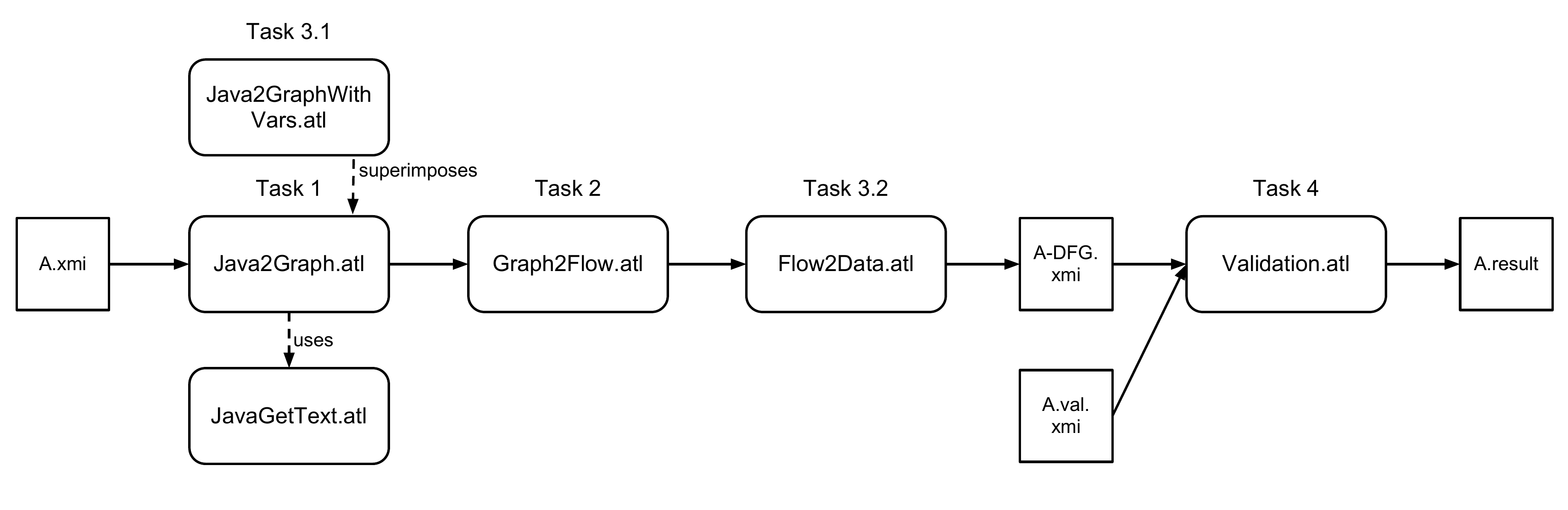}
	\caption{The chain of ATL transformations (intermediate models are omitted).}
	\label{fig:chain}
\end{figure}

The ATL solution for the Flowgraph Case Study is composed by a chain of ATL transformations orchestrated by an ANT file. The structure of the chain is illustrated in Fig. \ref{fig:chain} and detailed in the following.

\subsection{Task 1: Structure Graph}

The main transformation in Task 1 is Java2Graph.atl that implements a simple mapping between elements in the JaMoPP and FlowGraph metamodels. The mapping is illustrated in Table \ref{tab:mapping}. Each line of the table is encoded as a simple ATL rule. For instance in Listing \ref{lst:java2graph} we show the rule that translates while loops. Rules define model elements to match in the source model (\emph{WhileLoop}), model elements to generate in the target (\emph{Loop}), and values to assign to target properties. The rule in Listing \ref{lst:java2graph}  states that the \emph{expr} and \emph{body} references have to be filled with the result of the translation of the \emph{condition} and \emph{statement} of the matched element \emph{s}. In each of the rules of Java2Graph.atl a \emph{txt} attribute is filled with the concrete textual syntaxt of the element, calculated by calling a \emph{getText} attribute helper. The \emph{getText} helpers are defined in an ATL library, JavaGetText.atl, that is referenced by Java2Graph.atl.

\begin{table}[!h]
\begin{center}
\begin{tabular}{|p{5cm} | p{5cm}|}
\hline
{\bf Java Entities} & {\bf Flow Entities} \\
\hline
\multicolumn{2}{|c|}{\bf Methods}\\
\hline
ClassMethod & Method, Exit\\
\hline
\multicolumn{2}{|c|}{\bf Statements}\\
\hline
Block & Block\\
\hline
Condition & If\\
\hline
Return & Return\\
\hline
WhileLoop & Loop\\
\hline
Jump & JumpStmt\\
\hline
JumpLabel & Label\\
\hline
Continue & Continue\\
\hline
Break & Break\\
\hline
Other statements & SimpleStmt\\
\hline
\multicolumn{2}{|c|}{\bf Expressions}\\
\hline
EqualityExpression & Expr\\
\hline
RelationExpression & Expr\\
\hline
\end{tabular}
\end{center}
\caption{Java2Graph mapping}
\label{tab:mapping}
\end{table}

\begin{lstlisting}[language=ATL, style=AMMA, numbers=none, caption={A rule from Java2Graph.atl.}, label=lst:java2graph] 
rule WhileLoop2Loop {
	from
		s : JAVA!WhileLoop
	to
		t : GRP!Loop (
			expr <- s.condition,
			body <- s.statement,
			txt <- s.getText
		)
}
\end{lstlisting}

The library JavaGetText.atl contains a set of \emph{getText} attribute helpers, one for each metamodel element, that implement the model-to-text transformation task of the case study. In Listing \ref{lst:java-get-text} we show an excerpt of JavaGetText.atl, to illustrate its structure. Each helper is an OCL expression on the source model and the helpers call each other to construct complex concrete syntaxes. The excerpt in Listing \ref{lst:java-get-text} contains the necessary code to compute the textual syntax of an assignment of the form \emph{a=1;}.

\begin{lstlisting}[language=ATL, style=AMMA, numbers=none, caption={The model-to-text transformation JavaGetText.atl (excerpt).}, label=lst:java-get-text] 
helper context JAVA!ExpressionStatement def : getText : String =
	self.expression.getText + ';';
	
helper context JAVA!AssignmentExpression def : getText : String =
	self.child.getText + ' ' + self.assignmentOperator.getText + ' ' 
		+ self.value.getText;
		
helper context JAVA!LocalVariable def : getText : String =
	self.name;
	
helper context JAVA!DecimalIntegerLiteral def : getText : String =
	self.decimalValue;
\end{lstlisting}

\subsection{Task 2: Control Flow Graph}

Task 2 is implemented in the transformation Graph2Flow.atl. The transformation uses the \emph{refining mode} of ATL, allowing the developer to specify only the refinement part. In this case a set of rules adds the \emph{cfNext} reference that encodes control flow edges. All these rules have the structure shown in Listing  \ref{lst:graph2flow}: elements are matched and a \emph{cfNext} reference is added by calling a \emph{getNext} OCL helper. The logic for deriving control flow edges, detailed in the case study description, is encoded in the set of OCL \emph{getNext} helpers.

\begin{lstlisting}[language=ATL, style=AMMA, numbers=none, caption={A rule from Graph2Flow.atl}, label=lst:graph2flow] 
rule SimpleStmt {
	from
		s : GRP!SimpleStmt
	to
		t : GRP!SimpleStmt (
			cfNext <- s.getNext
		)
}
\end{lstlisting}

\subsection{Task 3: Data Flow Graph}
\subsubsection{Subtask 3.1} The construction of the data flow links requires to keep information, through the whole transformation chain, about variable uses and definitions. For this reason, the transformation in Task 1 has to be extended to avoid the loss of this information. We use the \emph{superimposition} mechanism to extend the Java2Graph.atl transformation in Task 1 with a set of additional rules and helpers. The rules of the superimposed transformation, Java2GraphWithVars.atl are executed together with the rules of Java2Graph.atl by the ATL virtual machine. Rules with the same name are overridden by the superimposed transformation (but this case does not apply to our scenario). Listing \ref{lst:java2graph-with-vars} contains the only two rules of Java2GraphWithVars.atl, that respectively create variables and parameters. A set of OCL helpers are called by \emph{getDefiners} and \emph{getUsers} to fill the definition and usage references. The set of helpers find uses and definitions by analyzing the position of the variable reference in the program tree. For instance a variable definition is detected whenever the variable reference \emph{isInLeftInAssignment} or \emph{isInUnaryModificationExpression}.

\begin{lstlisting}[language=ATL, style=AMMA, numbers=none, caption={Rules from Java2GraphWithVars.atl}, label=lst:java2graph-with-vars] 
rule LocalVariableStatement2Var {
	from
		s : JAVA!LocalVariable
	to
		t : GRP!Var (
			txt <- s.getText,
			definers <- Sequence{s.getLocalVariableStatement}->
				union(s.getDefiners),
			users <- s.getUsers																	
		)
}

rule OrdinaryParameter2Var {
	from
		s : JAVA!OrdinaryParameter
	to
		t : GRP!Param (
			txt <- s.getText,
			definers <- Sequence{s.getMethod}->union(s.getDefiners),
			users <- s.getUsers						
		)
}
\end{lstlisting}

\subsubsection{Subtask 3.2}

For the generation of data-flow links we implemented a variation of the algorithm in \cite{DRAGON} as a set of OCL helpers. The resulting iterative algorithm calculates for each flow instruction the set of definitions that the program needs when arriving to that point. It proceeds backwards by starting from variable uses, analyzing the successors of each flow instruction and propagating back the need for definitions.

\subsection{Task 4: Validation}
\begin{figure}[t]
	\center
	\includegraphics[width=8cm]{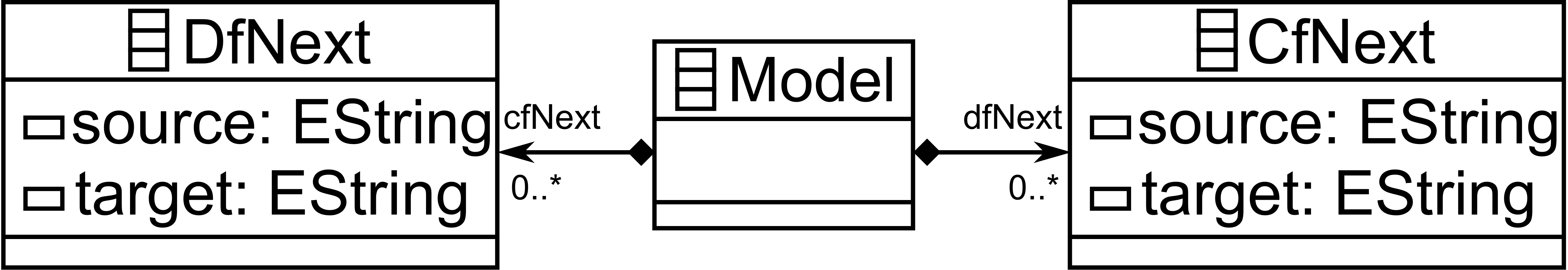}
	\caption{Metamodel for the Validation DSL.}
	\label{fig:validation}
\end{figure}

We implemented the validation task of the case study by an ATL model-to-text transformation (Validation.atl) that takes two models as input: the program dependence model generated by Task 3 and a user-provided specification model (Fig. \ref{fig:validation}). A set of OCL helpers iterate on the specifications and check that the correspondent dependency exists in the model. Vice versa, they also iterate on the dependency models to check that all the dependencies belong to the specification file. A textual list of missing links and false links is generated in output.

\section{Conclusion} \label{sec:conclusions}

The case study shows the applicability of ATL to complex transformation scenarios in program analysis. Table \ref{tab:size} presents size information on the implemented transformations\footnote{Performance evaluation of the proposed ATL solution can be found at \url{http://docatlanmod.emn.fr/TTC/Result.pdf}}. The transformations, beside intrinsic algorithmic complexity, look fairly readable.

\begin{table}[!h]
\begin{center}
\begin{tabular}{|p{3.5cm} | p{2cm} | p{2cm} | p{2cm}| p{2cm}|}
\hline
{\bf Transformation} & {\bf Task} & {\bf LOC} & {\bf Rules} & {\bf Helpers}\\
\hline
JavaGetText & 1 & 214 & 0 & 60\\
\hline
Java2Graph & 1 & 133 & 12 & 0\\
\hline
Java2GraphWithVars & 3.1 & 183 & 2 & 19\\
\hline
Graph2Flow & 2 & 324 & 7 & 28\\
\hline
Flow2Data & 3.2 & 92 & 2 & 6\\
\hline
Validation & 4 & 59 & 0 & 9\\
\hline
\end{tabular}
\end{center}
\caption{Transformation size}
\label{tab:size}
\end{table}

The problem can be modularized in a transformation network and concisely represented by using exclusively declarative transformation rules and helpers. All the phases are handled by the same transformation language: model-to-model, model refinement, model-to-text, validation. The case study represents an interesting illustration of the ATL application space.

\bibliographystyle{eptcs}
\bibliography{TTC2013ATL}

\end{document}